# Aromatic molecular junctions between graphene sheets: a Molecular Dynamics screening for enhanced thermal conductance

Alessandro Di Pierro [a], Maria Mar Bernal [a], Diego Martinez [a], Bohayra Mortazavi [b], Guido Saracco [a], Alberto Fina [a, *]

a- Dipartimento di Scienza Applicata e Tecnologia, Politecnico di Torino, Alessandria Campus,

Viale Teresa Michel 5, 15121 Alessandria, Italy

b- Institute of Structural Mechanics, Bauhaus-Universität Weimar

Marienstraße 15, D-99423 Weimar, Germany

## 1. Abstract

*The proper design and synthesis of molecular junctions for the purpose of establishing percolative networks of conductive nanoparticles represent an opportunity to develop more efficient thermally-conductive nanocomposites, with several potential applications in heat management. In this work, theoretical classical molecular dynamics simulations were conducted to design and evaluate thermal conductance of various molecules serving as thermal bridges between graphene nanosheets. A wide range of molecular junctions was studied, with a focus on the chemical structures that are viable to synthesize at laboratory scale. Thermal conductances were correlated with the length and mechanical stiffness of the chemical junctions. The simulated tensile deformation of the molecular junction revealed that the mechanical response is very sensitive to small differences in the chemical structure. The analysis of the vibrational density of states provided insights into the interfacial vibrational properties. A knowledge-driven design of the molecular junction structures is proposed, aiming at controlling interfacial thermal transport in nanomaterials. This approach may allow for the design of more efficient heat management in nanodevices, including flexible heat spreaders, bulk heat exchangers and heat storage devices.*

## 2. Introduction

Thermal management is central in several fast-growing technologies, from semiconductors industry[1], to avionics[2] and rechargeable lithium-ion batteries[3]. Inappropriate thermal management can result in overheating and consequent shortening of service life[4] and may even lead to hazardous conditions such as fire[5]. Aluminum and copper heat exchangers are common add-on solutions for high-drain integrated circuits, but they increase the number of components, system weight and assembly time, and therefore cost. Thermally-conductive chip packages made out of polymeric composites may represent a considerable improvement, but satisfactory materials are still lacking, due to insufficient thermal conductivity. In fact, research efforts are currently focused on the development of polymer composites and nanocomposites for efficient heat transfer[6-8]. Carbon nanotubes (CNT) and a variety of 2D materials as graphene-related materials (GRM)[9] , boron nitride[10] , fosforene[11], silicone[12] and recently borophene[13], attracted great research interest due to the outstanding properties evaluated through *in silico* approaches. In particular, thermal properties of graphene were studied in details and strictly correlated to its phonon transport properties[14-16]. Indeed, graphene was recognized as an excellent candidate for applications in thermally conductive materials. However, thermal transport in (nano)composites is strongly limited by the thermal resistances occurring between conductive particles in direct contact, as well as between particles and the surrounding matrix[8, 17]. The thermal resistance is caused by phonon scattering in the presence of structural defects and discontinuity[18], as well as acoustic mismatch[19] phenomena between the two components of the interface. In particular, the inefficiency of phonon transfer within a network of conductive particles in contact has been related to the “soft” interface, which does not allow transfer of efficient vibrational modes of phonons[7, 20]. A possible approach to decrease this contact resistance is by grafting molecular junctions[21] between nanoparticles, to increase the contact stiffness and enhance phonon transfer[22-23]. However, controlled chemical functionalization of graphene remains challenging and presents drawbacks, as covalent bonding on graphene induces rehybridization[24], acting as lattice defects[25] and altering the in-plane phonon modes[26], ultimately affecting the thermal properties of graphene[27-28]. On the other hand, non-covalent functionalization based on secondary interactions[29-30] is known to preserve the hybridization of graphene but is expected to deliver rather limited reductions in contact resistance.

Non-Equilibrium Molecular Dynamics (NEMD) is a well-established numerical method to investigate thermal conductivity and interfacial thermal properties[31]. The Interfacial Thermal Conductance (ITC), defined in this paper as the conductance between solid particles and the polymer matrix, reflects the physico-chemical properties of the interface[6, 20]; in fact, chemical compatibilization through organic chain grafting onto the solid particles was typically found to be effective to improve thermal transport[32-34]. Wang et al[33] simulated covalently alkyl functionalized graphene platelets inside a polyethylene matrix, showing that an higher amount of linkers, and longer protruding linkers in matrix increases ITC. The Vibrational Density Of States (VDOS) highlighted that functionalization induced a shift in frequencies where phonon match occurs more between polymer and fillers. A similar layout was reported by Wang et al.[34], via covalent functionalization of graphene with a variety of small dangling molecules: in this case, the VDOS spectra of butyl functionalized graphene found in the best phonon match with paraffin polymer matrix, resulting in a remarkable ITC increase. Later, Shen and coworkers[35] investigated the ITC in epoxy resin/graphene nanocomposites, as a function of the chemical functionalization of graphene, and showed that the highest ITC reduction occurs with triethylenetetramine moieties, because of their ability to penetrate in the epoxy resin and form covalent bonds. On the other hand, the thermal conductance associated with filler-filler contacts, in this work referred to as Thermal Boundary Conductance (TBC), was also studied numerically by various authors, specifically between carbon nanotubes and graphene platelets surface-functionalized with small bridging groups such as oxygen or methylene bridges[36-39], short alkyl chains[40] or benzene[41]. TBC of alkyl molecular junctions covalently bound between two graphene flakes was investigated in details by Li and coworkers[40] via Density Functional Theory (DFT), showing a correlation between TBC and the stretching imposed to the molecular junction. Recently, we also reported NEMD simulations of edge-to-edge alkyl junctions between graphene nanoribbons[42], where covalent bonding was compared to Van der Waals forces between interpenetrating molecules.

While the above mentioned numerical simulations focused on small chemical moieties or alkyl chains as bridging agents between nanoparticles, the synthesis of these types of junctions between carbon nanoparticles is experimentally very challenging. Therefore, more viable chemical structures need to be addressed to allow for practical applications. Li *et al.*[30] addressed for the first time aromatic molecules as thermal junctions in π-stacked graphene sheets, and compared it with a para-disubstituted benzene covalently bound junction. Han *et al.*[43] reported aminosilane functionalization between a silicon substrate and GRM for heat management in electronic hotspots achieving lower operating temperatures than unfunctionalized samples. Furthermore, we recently reported the experimental and computational study of graphene nanoplatelets edge-functionalized with 1,5-bis(4-aminophenyloxy)pentane or 4-aminophenol to produce molecular junctions based on covalent bonding or as secondary interactions, respectively[44].

Building on previous work, the present contribution addresses the TBC calculation of various aromatic molecules as possible candidates for the use in thermal molecular junctions, focusing on chemically viable systems that are synthesizable at laboratory scale. Different edge-grafted molecules, which act as thermal bridges between graphene nanosheets, are evaluated with respect to TBC via full atomistic NEMD. The contribution of vibrational modes, molecule stiffness, chain chemistry and segments lengths were systematically analyzed, providing a computational screening for molecules to be utilized experimentally.

## 3. Molecular Dynamics modeling

NEMD calculations were carried out with Graphics Processing Unit (GPU) accelerated[45] Large-scale Atomistic Molecular Massively Parallel Simulator (LAMMPS) code. The COMPASS force field developed by Sun[46] was chosen for the introduction of the atomic interactions in this work. COMPASS is a widely used force field in the molecular dynamics simulation of organic materials, including GRM polymer nanocomposite[35, 47]. Although the Tersoff potential is more frequently used to calculate graphene thermal conductivity, COMPASS was also used by Zhang et al[48] to calculate a value of 550 $Wm^{-1}K^{-1}$ for thermal conductivity in graphene finite slab. COMPASS includes a much larger chemical species database than Tersoff and AIREBO, allowing simulating organic chemical structures as well as non-bond interactions in the condensed phase. It also contains anharmonic bonding terms, providing a more accurate representation of thermal transport processes[49], which appear relevant in molecular junctions[50]. In this work, non-bond interactions were defined by the Van der Waals contribution from the COMPASS built-in Lennard Jones 9-6 function and the coulomb electrostatic term, by the use of atomic partial charges[46], with a 10 Å cut-off. Velocity Verlet algorithm recalculated positions and velocities every 0.25 fs while the initial partial charges assignment and equilibration was set through Qeq algorithm[51]. Periodic Boundary Conditions (PBC) were set in x,y,z. However, being a supercell system, the simulation domain was padded to fit the model and empty space was set to more than double of the cut-off distance in order to avoid PBC interaction.

The model was composed of about 4100 atoms, divided in two graphene nanoribbons (approximately 100 Å by 50 Å) grafted through the armchair edge by six parallel molecular junctions. Because the model size is shorter than the phonon mean free path in graphene, no inner flake phonon scattering flake is expected[16]. This edge-to-edge layout is depicted in Figure 1A, with six 1,5-bis(p-phenylenoxy)pentane junctions as an example. All aromatic molecules ends were bonded in para (p-) with the graphene nanoribbons (Figure 1B') except for the fully aromatic acene-based thermal bridges (Figure 1B''). The molecular junction length refers to the distance "d" between platelets edges (Figure 1B). All the distances reported herein were determined through VMD[52] tools from time-averaged measurements of the six junctions. Furthermore, the C-C distance was found 1.39 Å, in fair agreement with literature values (1.40 Å) for other force fields commonly adopted in thermal investigations with graphene[53].

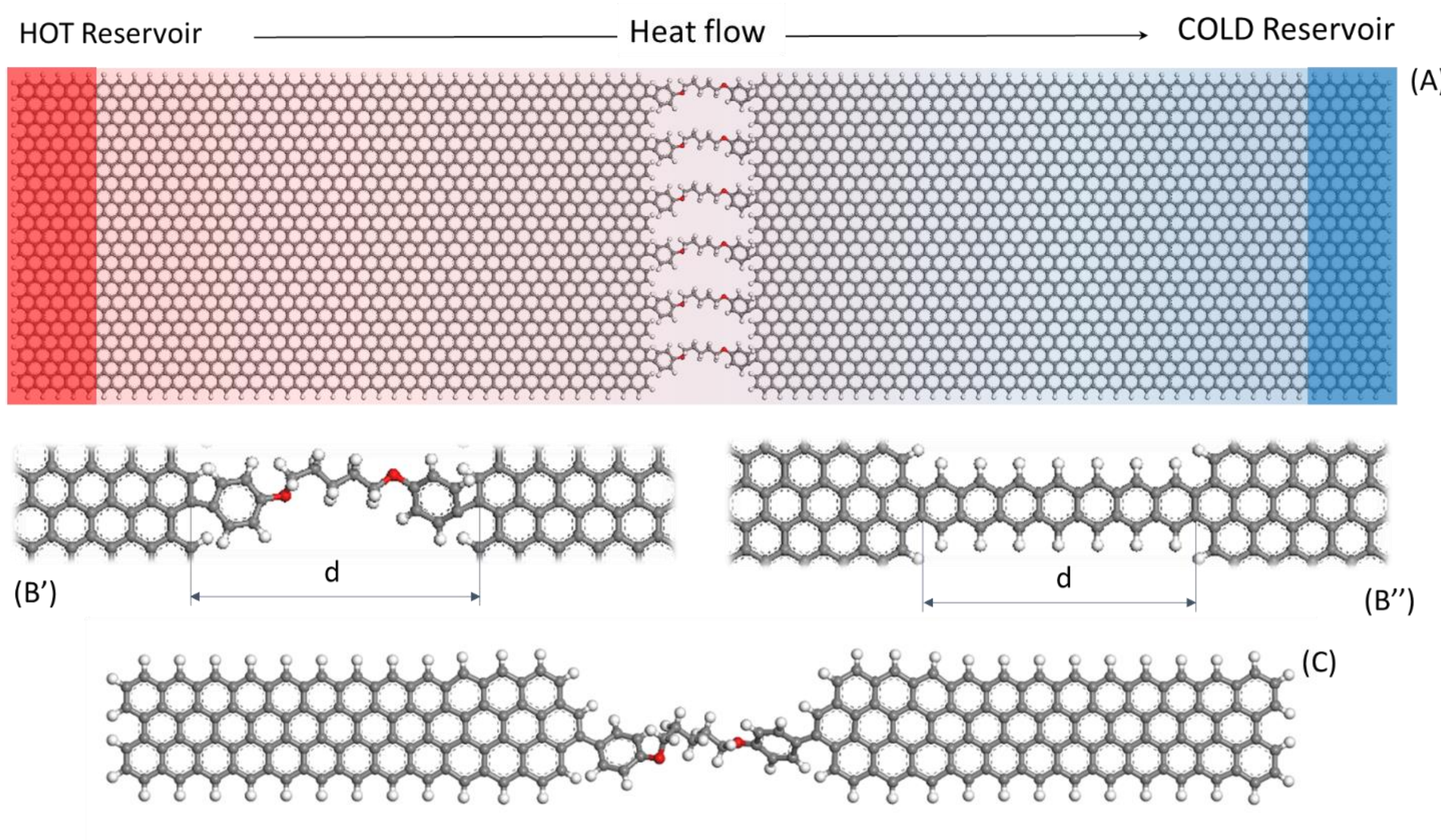

**Figure 1.** (A) Planar model representation and NEMD layout, (B') junctions detail of 1,5-bis(p-phenylenoxy) pentane, C5OP, and (B'') Heptacene, with dimensioning ("d") for platelets distance. (C) The model for tensile testing with C5OP, as example molecule.

The simulation procedure followed a well-established scheme: the whole system was initially relaxed through an equilibration period of 125ps in NVT (constant Number of atoms, Volume and Temperature) canonical ensemble at 300K, within a temperature range in which the quantum effect was previously reported to be negligible[54]. At the end of the thermal equilibration, followed 250ps of thermo-stated preheating with thermostats. In this stage, Nosé-Hoover thermostats, were set to 310K and 290Kand applied to the model ends, about 8 Å along the x coordinate. Here, all the atoms in the system, except those in thermo-stated regions, run in NVE (constant Number of atoms, Volume and Energy) and a thermal gradient is gradually established inside the model. With the system in steady state, for 4.8 ns, data collection[27, 55] of energy and temperature was performed. The thermal flow through thermostats was obtained from the slope the energy versus time plot[56], while the group temperatures were computed from the averaging the instantaneous local kinetic temperature[27, 55] as reported in Equation 2:

$$T_i(slab) = \frac{2}{3N_i k_B} \sum_j \frac{p_j^2}{2m_j} \qquad (2)$$

where $T_i$ (slab) is the temperature of $i^{th}$ slab, $N_i$ is the number of atoms in $i^{th}$ slab, $k_B$ is the Boltzmann's constant, $m_j$ and $p_j$ are atomic mass and momentum of atom $j$, respectively. With the system is in steady state, the quantity of heat being injected into the heat source and removed from the heat sink evolves with time linearly, equivalent slopes (Supporting Information, Figure S1), evidences the constant energy inside the ensemble and the steady heat flux which passes through the model[56]. The temperature profile was obtained by splitting the model length into 22 slabs and by time-averaging the temperature of each slab. In this calculation were excluded the regions where the temperature evolves not linearly, as close to the thermostats and across the junction. The molecular thermal conductance, $G_m$ in pW/K has finally calculated by the Fourier's equation (3).

$$G_m = \frac{q_x}{n \cdot \Delta T} \qquad (3)$$

Where $q_x$ [eV/ps] represents the heat flow, $n$ is the number of molecules bridging the nanoribbons and $\Delta T$ [K] is the calculated temperature as projection of the two linear fit from the temperature-length plot at junction edge. The calculated temperature profiles along the graphene nanoribbons are reported in Supporting Information (Figure S2). The thermal conductance values proposed in this work were calculated from different velocity seeds simulations, with a minimum of 3 replicas for each model.

To investigate the phonons in molecules and graphene[26], vibrational spectra has been calculated for the isolated molecules as well as for a hydrogen-terminated graphene nanoribbon. The VDOS were calculated through discrete Fourier transform of the x, y and z velocity autocorrelation function in equation 4:

$$D(\omega) = \int_0^{\tau} \langle v(0) \cdot v(t) \rangle e^{-i\omega t} dt \qquad (4)$$

where $D(\omega)$ is the VDOS at the frequency $\omega$, $\langle v(0) \cdot v(t) \rangle$ is the correlation function of atoms velocities. In VDOS run, the temperature was initially set in canonical ensemble at 300 K for 250 ps. Data collection was performed in microcanonical ensemble, saving the velocities every timestep for 12.5 ps. In this calculation, the PBC were set as fixed with about 50Å of padding room, determining no interaction at simulation box boundaries.

The Elastic modulus was calculated through linear fitting up to 10% of deformation in elastic regime. Stress-strain plots are reported in Figure S4. The result of this test should be considered only in relative terms between the molecules and not as absolute Young modulus. Uniaxial strain simulations were conducted for six replicas along z at 0.1K, similarly to metal nanowires[57-58], to minimize the velocity fluctuation noise in stress tensors calculation. For this study, a different specimen made out of two small graphene ribbons (about 80x10 Å$^2$) in single junction has been designed for every molecule (as example, C5OP in shown in Figure 1E). The simulation box PBC were set as fixed in x and periodic in y and z, adopting a 50 Å padding room as already adopted in VDOS calculation. An initial relaxation step in NVT was set bringing the temperature from 5K to 0.1K in 5ps. Then, the left slab of the specimen was fixed and a longitudinal uniaxial velocity of 0.5 Å/ps was applied to the right-end graphene ribbon generating a constant pulling force overall the system. The simulation run in NVE for 20ps or until the molecule broke, whichever occurred first.

## 4. Results and Discussion

Sixteen molecular junctions were considered in this study, as summarized in Table 1. Aliphatic/aromatic molecules represent the largest subset, and include a variety of diphenyloxyalkanes (A-G, in Table 1): diphenoxymethane (C1OP); 1,2-diphenoxyethane (C2OP); 1,3-diphenoxypropane (C3OP); 1,4-diphenoxybutane (C4OP); 1,5-diphenoxypentane (C5OP); 1,6-diphenoxyhexane (C6OP); 1,7-diphenoxyheptane (C7OP). 1,3 dibenzylbenzene (PCP) an 1,7 diphenylheptane (C7P) were also addressed as ether-free counterparts. Diphenoxybenzene (POP, H in Table 1) exploited a central aromatic moiety instead of the aliphatic one. Fully aromatic junctions include, biphenyl (BP), phenanthrene (PH) and pyrene (PY) (K-M in Table 1). It is worth mentioning that all of these junctions may in principle be obtained between graphene sheets via aryldiazonium chemistry, as demonstrated experimentally for C5OP[44]. For sake of comparison, junctions based on acenes containing 3 (anthracene, ACN), 5 (pentacene, PCN) or 7 (heptacene, HCN) benzene rings, were also simulated (N-P in Table 1). While their use as thermal junction remains theoretical owing to the difficulties in the synthesis of such molecular junctions, six-member carbon rings between graphene nanoribbons were previously addressed as a model system for graphitized polyimide/graphene composites[59]. It is worth noting that acene junctions preserves the conjugation of $sp^2$ carbon across the contact of the two graphene sheets; in fact acene-bridged graphene sheets could even be considered as a single graphene with rectangular holes in its structure, that was previously addressed by Yarifard et al.[60]

As a first case study, C5OP was analyzed in a range of parameters, including Thermostat Temperature Difference (TTD), molecular conformation and junction topology. The TTD effect was evaluated keeping the simulation equilibrium temperature centered at 300K and increasing the TTD from 20K to 25K and 30K, leading to $G_m$ values of 139±7, 131±6 and 135±4 pW/K, respectively, which are considered constant within the experimental uncertainty. When shortening the alkyl chain in aliphatic/aromatic (CnOP) junctions to C4OP and C3OP, an enhancement of $G_m$ to 148±23 and 161±20 pW/K was obtained, respectively. Further shortening to C2OP and C1OP exhibited further increased values up to 182±5 and 208±53 pW/K, thus confirming the increasing conductance with shortening junction length. On the other hand, as expected, longer component of the alkyl portion reported lower $G_m$: 130±9 pW/K in C6OP and 123±6 pW/K in C7OP. It is also worth noting that the calculated thermal conductances of those molecules were in the same order of magnitude of similar length[40, 42] alkyl ones.

To investigate the CnOP molecular junctions stiffness as a function of the chain length, a set of tensile simulations have been performed. Resulting elastic moduli (E) were found in the range between 145 and 53 GPa for C1OP and C6OP, respectively, with a decreasing trend over values with increasing alkyl chain length (Table 1). As matter of comparison, elastic modulus of 487 GPa was calculated for pristine graphene slab. Despite this value is significantly lower, compared to 980 GPa previously reported[27], (which may be related to the different force field as well as to the small model size), a comparison of the obtained modulus remains possible within this set of results. Interestingly, the tensile modulus trend vs. CnOP junction length was found to be consistent with the decrease in thermal conductance (Figure 2A), despite an odd-even effect is observed in calculated elastic modulus. To gain further insight about the role of chain length, VDOS were calculated for the different junctions (Figure 2B). Graphene exhibits a broad region of vibrational states up to 5 THz, a sharp peak at 45.2 THz (called G peak from Raman analysis, usually at 47THz[15]) and broad peak around 17 THz. For molecular junctions, several additional bands are observable in the regions 0 to 25 THz and 42 to 46 THz. However, very limited match is observed between VDOS spectra for pristine graphene and in the presence of CnOP molecular junctions. The almost linear decay in thermal conductance reported in Figure 2, combined with the phonon spectra mismatch, is indicative of a diffusive regime with scattering at the interface strongly limiting heat transfer.

(A)

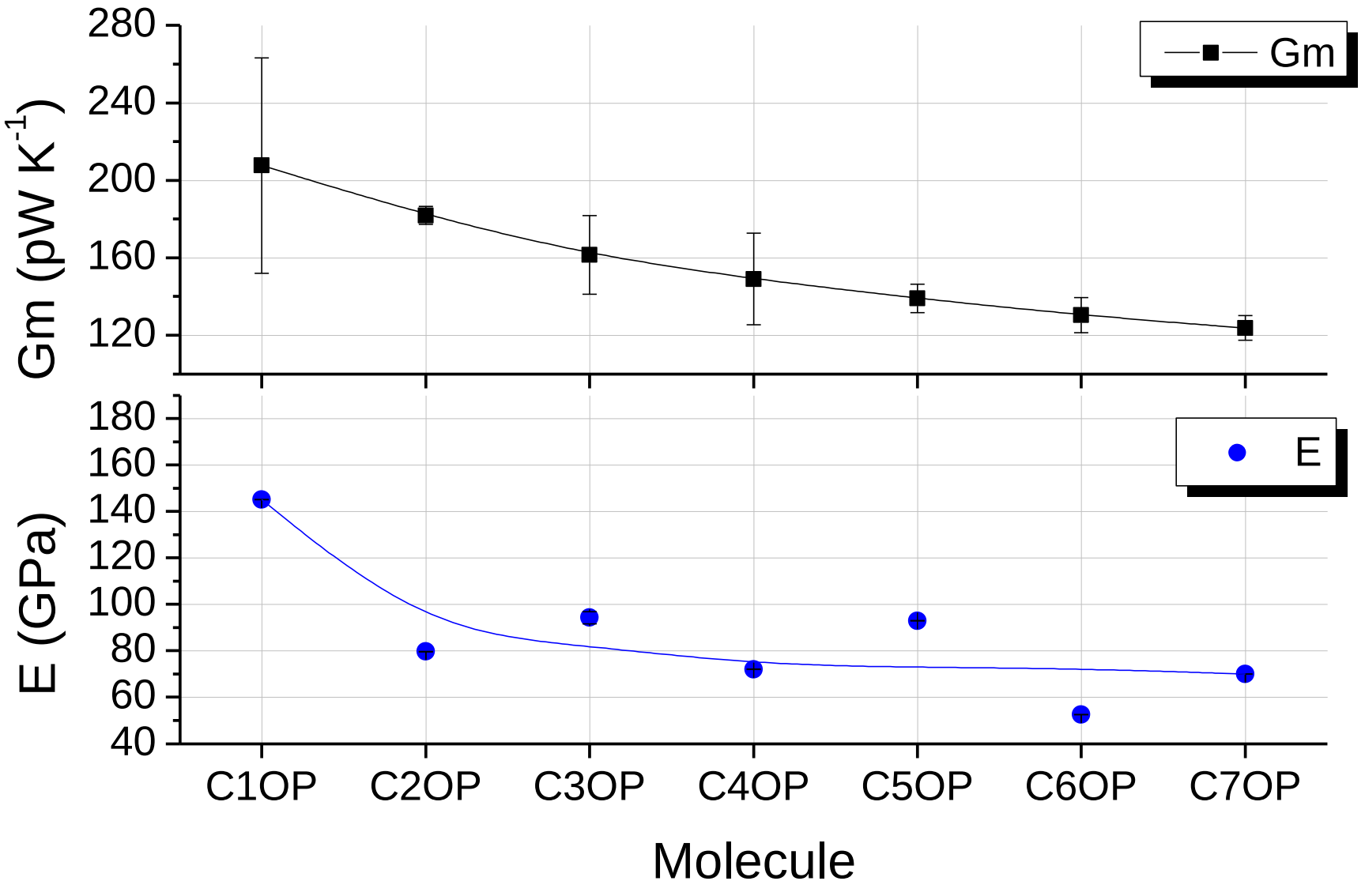

(B)

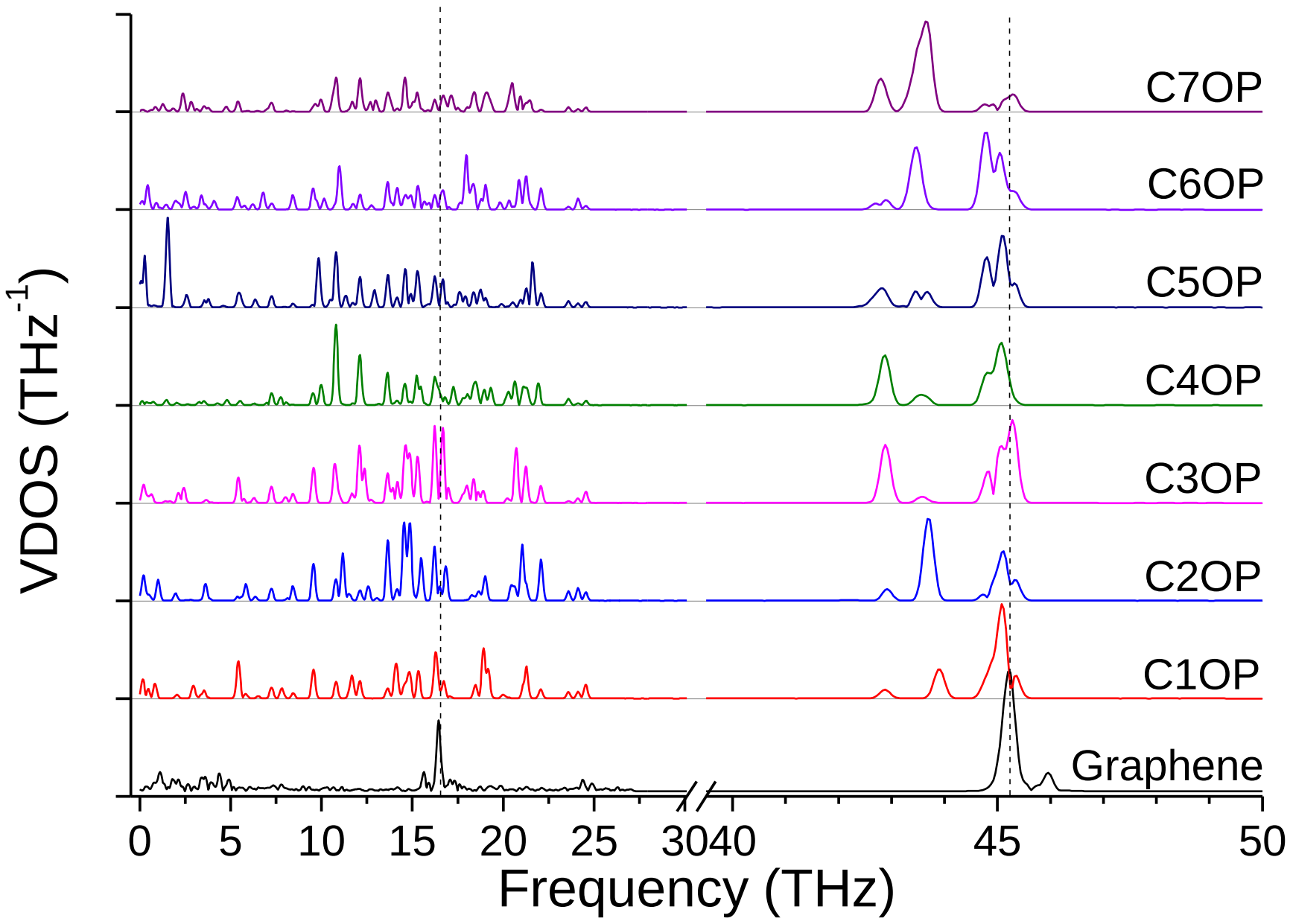

**Figure 2.** (A) $G_m$ and elastic modulus E for aliphatic-aromatic molecular junctions, the line guides the eye through values reported in Table 1. Error bars in E are below 1% and barely visible. (B) VDOS for the same aliphatic-aromatic molecular junctions, scale in 40-50 THz is magnified by a factor of three. The dashed lines highlight the most intense graphene peaks at 16.5 THz and 45.2 THz (G-peak).

The role of ether bond was studied comparing junctions with similar length, namely C5OP and C7P (E, J in Table 1). Despite the stiffness of C7P (65 GPa) was found approx. 30% lower than C5OP, the 139±33 pW/K conductance for C7P is equivalent to the value for C5OP. This finding suggests that the presence of the soft alkyl component in the center of the molecular junction limits the overall heat transfer and ether substitution in a relatively long alkyl chain has a negligible effect. The VDOS for C7P is reported in Figure 3, which shows the main contribution from alkyl moieties as broad peaks in the 42-44 THz region, in fair agreement with the spectrum for C5OP. The role of flexible groups (ether vs. methylene bridges) was further investigated in more rigid aromatic structure. Junctions based on diphenoxybenzene (POP, d=16.6 Å, H in Table 1) and dibenzylbenzene (PCP, d=16.3 Å, I in Table 1) were compared in both thermal conductance and stiffness. The thermal conductances for these systems were found about 260±16 pW/K for POP and 221±14 for PCP, suggesting ether bridges are more effective that methylene in phonon transfer. Furthermore, these values were compared to a junction of comparable length, C4OP, (148±23 pW/K, d = 16.3 Å, D in Table 1) which clearly evidenced the advantage of aromatic structures in terms of thermal conductance. The analysis of VDOS spectra (Figure 3) provided further support for the differences in heat transfer. Indeed, signal in the region of 43-44 THz in VDOS spectra progressively reduced from C7P to PCP, while no significant counts in this band remains for POP. Furthermore, a stronger peak rises around 16.5 THz in PCP and POP, while the G peak for POP junctions appears to best overlap with the main vibration of pristine graphene.

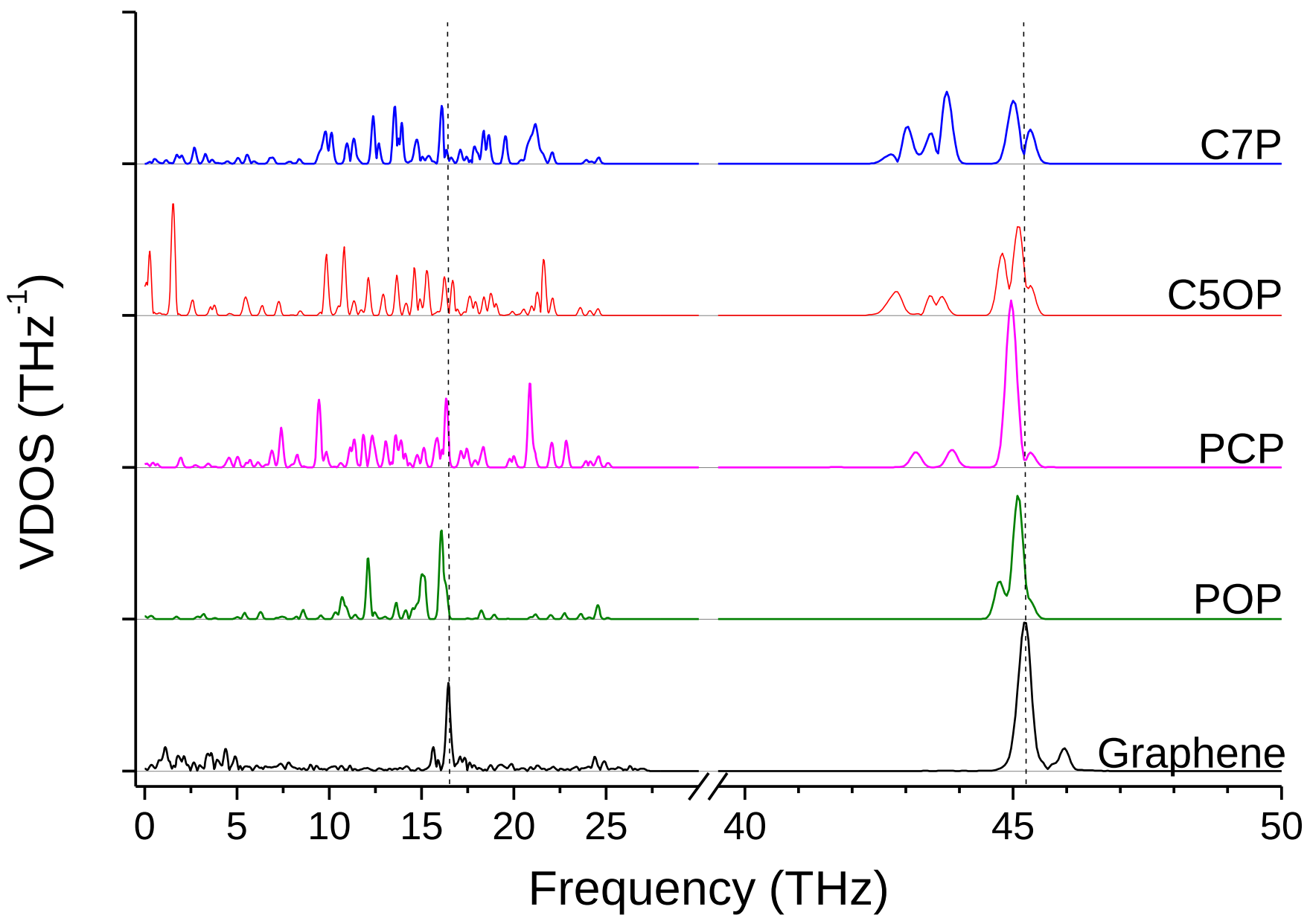

**Figure 3.** VDOS for molecular junctions with/without ether bridge, the scale in 40-50 THz is magnified by a factor of three for clarity. Dashed lines highlights the most intense graphene peaks at 16.5 THz and 45.2 THz (G-peak).

Tensile simulations highlighted a wide gap between elastic moduli for POP (152 GPa) and PCP (85 GPa). This difference suggested that the deformability of flexible moieties is indeed crucial in controlling the overall junction stiffness, also when rigid aromatic rings are present in the molecular structure. Indeed, carbon-oxygen bond parameters in COMPASS evidence for significantly higher stiffness compared to C-C (2$^{nd}$ 3$^{rd}$ and 4$^{th}$ power coefficients for c4o-o2e are remarkably higher than c4-c4[46], details in Supporting Information, Table S1 ad Figure S4), directly affecting the vibrational states. Based on this landmark, polyaromatic hydrocarbons were also studied as potentially effective molecular junctions. In fact, polyaromatic molecules exhibits highly delocalized electronic structures, and limited conformational freedom of the molecule, becoming progressively more similar to graphene itself when increasing the number of condensed aromatic rings. Indeed, VDOS analysis of selected polyaromatic junctions (Figure 4) confirmed the general excellent overlapping of the two most intense signal for polyaromatic junction with the 16.5 THz and 45.2 THz peaks for graphene. Some differences are clearly observable between the aromatic junctions and graphene, the molecular junctions own more vibrational modes in the region 5-20 THz. A thermal conductance of 472±26 pW/K was calculated for biphenyl junction (BP, K in Table 1) and further enhancement was obtained for higher aromatic condensation, as 608±25 pW/K was calculated for phenanthrene (PH, L in Table 1) and 648±14 pW/K for pyrene (PY, M in Table 1). The tensile deformation of aromatic junctions reflects a higher stiffness than the alkyl based ones, with calculated elastic moduli of 141 GPa for biphenyl, 165 GPa for phenanthrene and 179 GPa for pyrene.

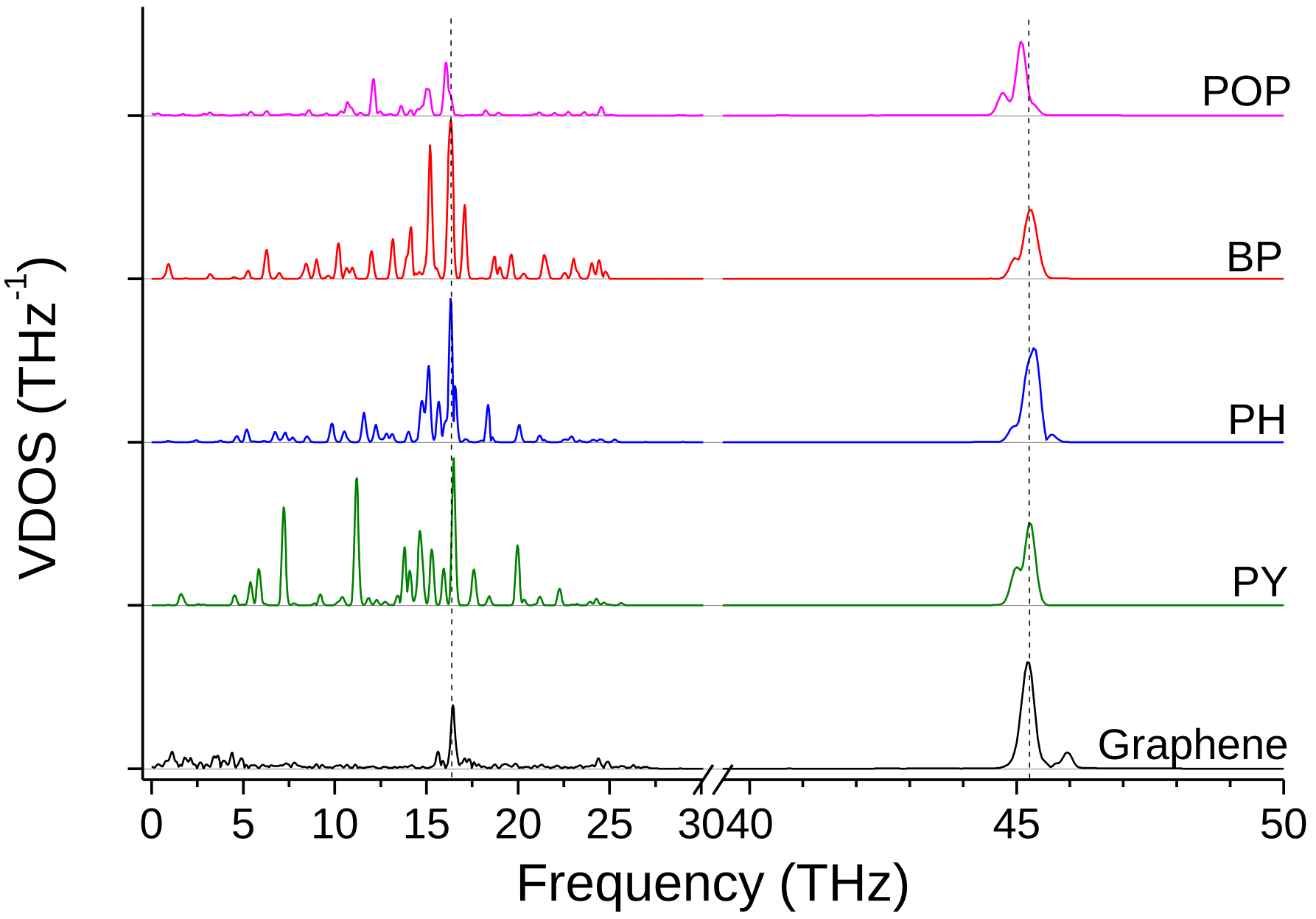

**Figure 4.** VDOS for aromatic molecules, scale in 40-50 THz is magnified by a factor of three. Dashed lines highlights the most intense graphene peaks at 16.5 THz and 45.2 THz (G-peak).

Based on the results obtained for aliphatic and aromatic junctions described so far, the combination of length and stiffness of the bridging chain appear to control the overall efficiency of the thermal transport through the interface. In fact, short and stiff junctions lead to the highest values of thermal conductance, as depicted in Figure 5.

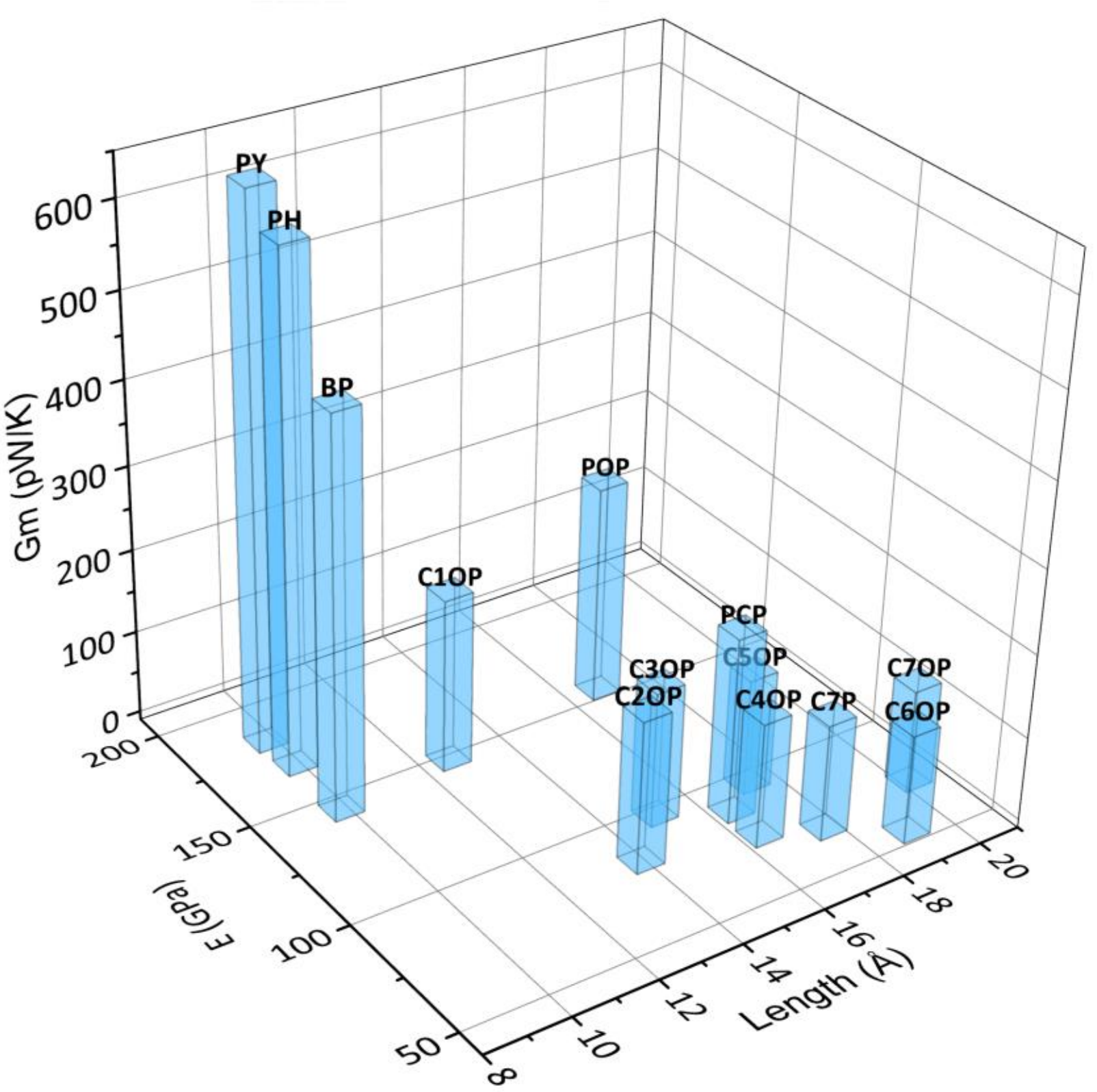

**Figure 5.** Calculated thermal conductance $G_m$ of thermal junctions as a function of length and stiffness.

Finally, acene-based junctions made of 3, 5 and 7 aromatic rings (anthracene, ACN, N in Table 1; pentacene, PCN, O in Table 1; heptacene, HCN, P in Table 1) were addressed. Despite the exploitation of acenes in molecular junctions is extremely challenging, these junctions were addressed here as a kind of theoretical "upper bound" for the thermal conductance for thermal bridges between graphene sheets, as it preserves the conjugation of $sp^2$ carbon across the contact of the two graphene sheets. The tensile testing revealed almost constant elastic modulus values, namely 279 GPa for ACN, 285 GPa for PCN and 268 GPa for HCN. On the other hand, $G_m$ values of 1121±35, 1076±44 and 1007±40 pW/K were obtained for ACN, PCN and HCN, respectively. These values appear to be only slightly affected by the acene length, which more than doubles, rising from 7.4 Å for ACN to 17.1 Å for HCN. These results suggest the phonon transfer on these junctions to be ballistic, in contrast with the diffusive regime observed in alkyl junctions. As expected, the VDOS (Figure 6) for polyaromatic junctions closely matches the main bands for pristine graphene, similarly to other aromatic junctions (Figure 4).

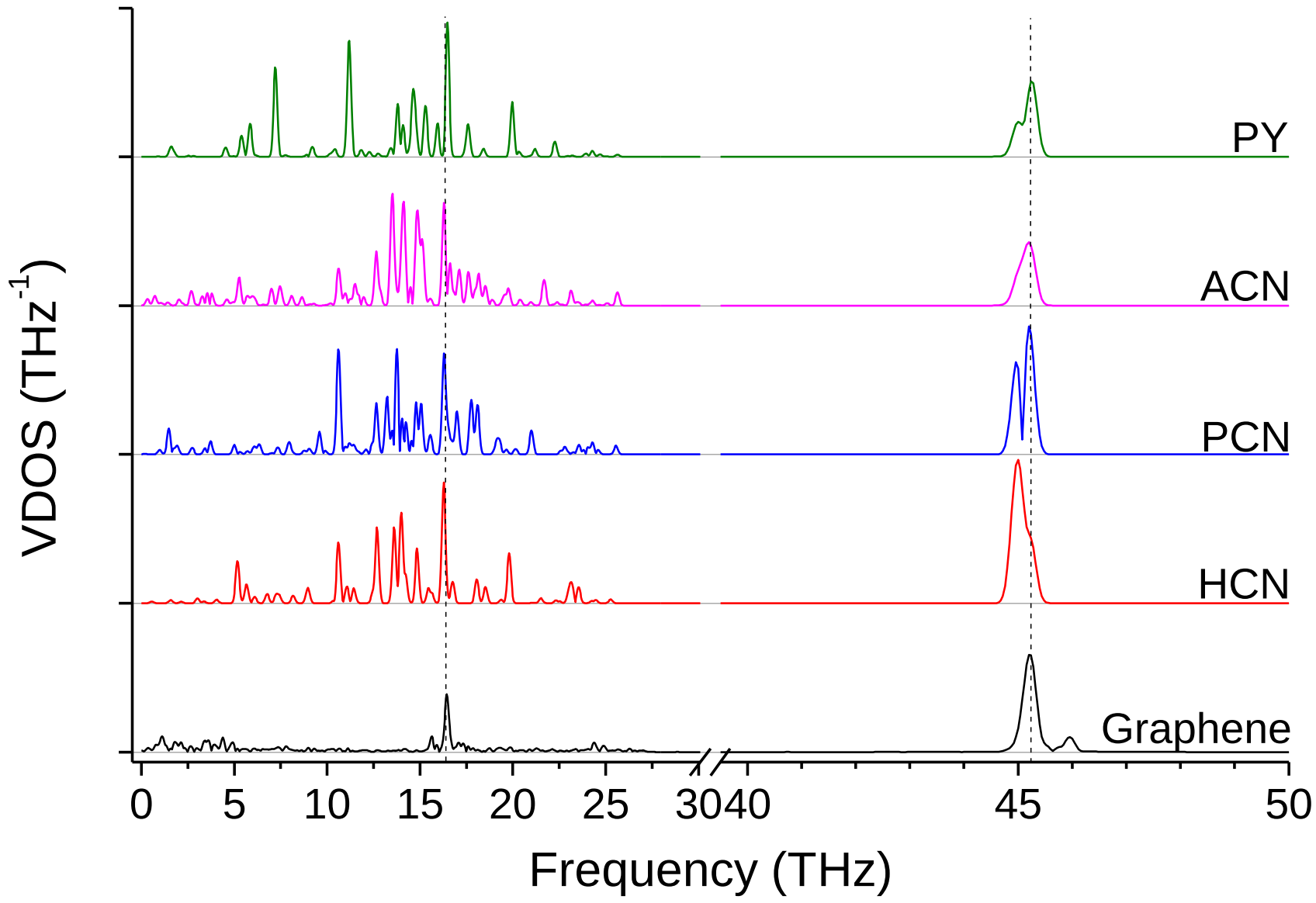

**Figure 6.** VDOS in acenes, scale in 40-50 THz is magnified by a factor of three. Dashed lines highlights the most intense graphene peaks at 16.5 THz and 45.2 THz (G-peak). Pyrene is added as a matter of comparison.

Table 1. The molecular junctions addressed in this work: Chemical structures, molecular thermal conductance ($G_m$), elastic modulus (E) and application distance (d). All molecules from A to M are bonded in para (p-) except of acenes (N-P) which exploits aromatic bonding.

| # | Long name | Short name | $G_m$ [pW/K] | E [GPa] | d [Å] | Structure |
|---|---|---|---|---|---|---|
| A | diphenoxymethane | C1OP | 208±53 | 145 | 12.5 | 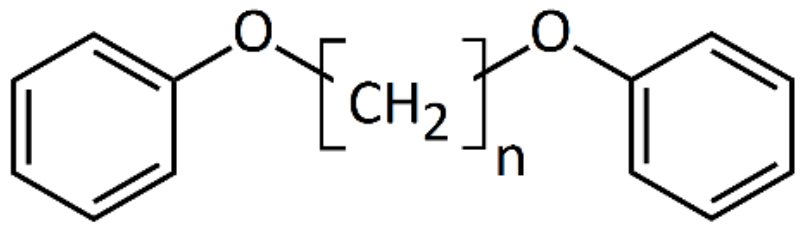 1,n-diphenoxy"alkyl" (CnOP) |
| B | 1,2-diphenoxyethane | C2OP | 182±5 | 80 | 13.8 | |
| C | 1,3-diphenoxypropane | C3OP | 161±20 | 94 | 14.9 | |
| D | 1,4-diphenoxybutane | C4OP | 148±23 | 72 | 16.3 | |
| E | 1,5-diphenoxypentane | C5OP | 139±7 | 93 | 17.2 | |
| F | 1,6-diphenoxyhexane | C6OP | 130±9 | 53 | 18.8 | |
| G | 1,7-diphenoxyheptane | C7OP | 123±6 | 70 | 20.0 | |
| H | 1,3-diphenoxybenzene | POP | 260±16 | 152 | 16.6 | 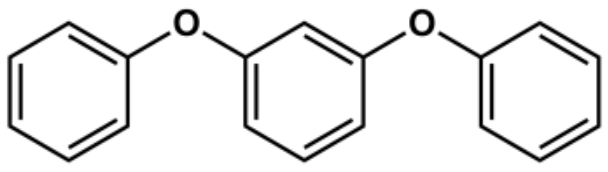 |
| I | 1,3-dibenzylbenzene | PCP | 221±14 | 85 | 16.3 | 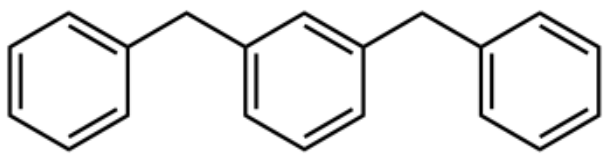 |
| J | 1,7-diphenylheptane | C7P | 139±33 | 65 | 17.5 | 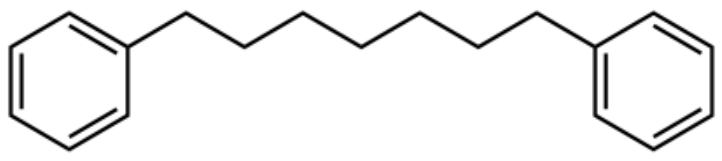 |
| K | Biphenyl | BP | 472±26 | 141 | 9.7 | 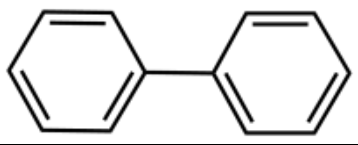 |
| L | Phenanthrene | PH | 608±25 | 165 | 9.8 | 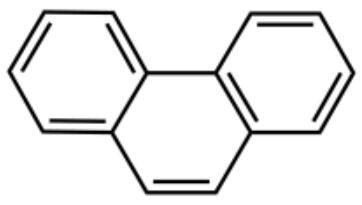 |
| M | Pyrene | PY | 648±14 | 179 | 9.7 | 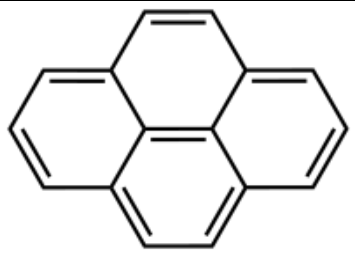 |
| N | Anthracene | ACN | 1121±35 | 279 | 7.4 | 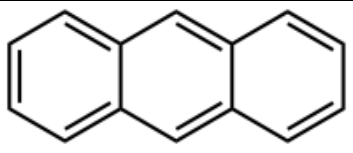 |
| O | Pentacene | PCN | 1076±44 | 285 | 12.3 | 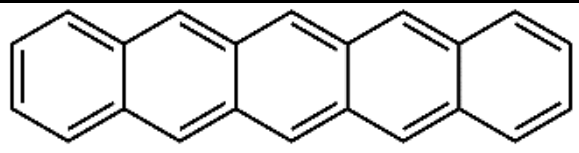 |
| P | Heptacene | HCN | 1007±40 | 268 | 17.1 | 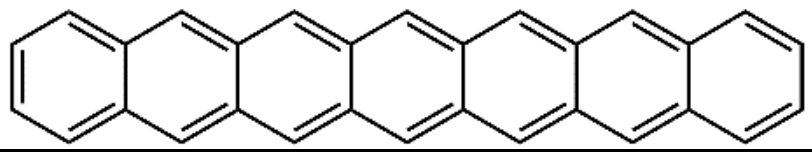 |

## 5. Conclusions

The thermal conductance of molecular junctions between graphene nanoribbons were investigated by means of non-equilibrium molecular dynamics simulations. Particular focus was dedicated to experimentally synthesizable junctions for the purpose of improving the thermal percolation within networks of graphene nanoflakes. Thermal boundary conductance was found to strongly correlate with molecular junction length and stiffness. Calculated thermal conductance was found to be lowest for aliphatic/aromatic junctions characterized by the long and flexible alkyl chain and highest for the short and rigid polyaromatic bridging molecules (e.g. phenanthrene and pyrene). Chemical structures that are able to change structural conformations due to the presence of free rotating single bonds, appeared to directly limit the thermal transport capability. Differences were also observed between junctions containing methylene and ether groups, related to the different features of their bonds to carbon.

In terms of phonon density of states, the overlapping of molecular junction VDOS with the spectra of graphene was found to correlate with the thermal conductance of the junctions. While VDOS for aliphatic/aromatic junctions exhibited limited overlapping with the main vibration bands of graphene at both 16.5 THz and 45.2 THz, aromatic and polyaromatic structures displayed vibrational spectra similar to graphene.

Finally, acene-based junctions, which preserve the conjugation of $sp^2$ carbon across the contact of two graphene sheets, were included in this study for sake of comparison. Their vibrational density of state was found comparable to graphene, resulting in values of thermal conductance that exceed 1000 pW/K. We found that thermal conductance varies little with acene length, suggesting ballistic phonon transfer trough the interface. While acene junctions synthesis is currently very challenging, the simulated conductance values obtained here should be considered as a theoretical upper limit to the thermal performance of any experimentally viable junction.

The screening of the different covalent junction structures addressed in this paper provides a guideline for the development of thermally efficient interfaces to be exploited in graphene-based thermal management materials. Furthermore, the molecular thermal conductance values calculated here can be used in upper-scale continuum models to predict effective thermal conductivity in graphene-based laminates or polymer nanocomposites. Proposed methodology can be also extended for the engineering and design of thermally percolative networks for the enhancement of heat transfer in other nanostructured systems.

## 6. Authors' contributions

A. Fina conceived the experiments, interpreted results and led the project, A. Di Pierro carried out MD simulations, post processing and interpreted results, Diego Martinez provided VDOS spectra interpretation. B. Mortazavi supported NEMD calculation and discussion of the results. Maria Mar Bernal supported chemical interpretation and G. Saracco participated to the discussion of results. Manuscript was mainly written by A. Di Pierro and A. Fina.

## Acknowledgements

This work has received funding from the European Research Council (ERC) under the European Union's Horizon 2020 research and innovation programme grant agreement 639495 — INTHERM — ERC-2014-STG. B.M. acknowledges the financial support by ERC for COMBAT project (Grant number 615132).

## Supporting information

Plots of NEMD energy, temperatures and stress-strain with tensile simulation details and force field parameters description.

## 7. Table of contents

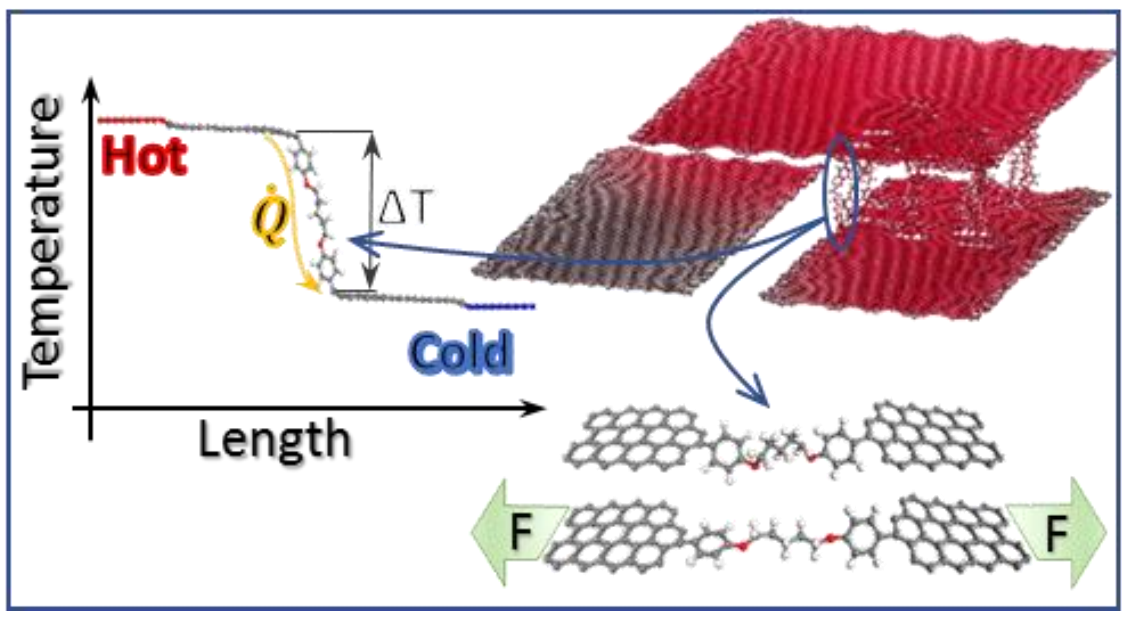

The tuning of covalently bound molecular junctions could increase heat transfer between graphene platelets inside nanocomposites materials.